\documentclass[letter,english,prl,twocolumn,showpacs]{revtex4}
\usepackage{ae} 
\usepackage[T1]{fontenc}
\usepackage[ansinew]{inputenc}
\usepackage{amsmath}
\usepackage{amssymb}
\usepackage{amsfonts}
\usepackage[dvips]{graphicx}
\usepackage{ifthen}
\usepackage{eurosym}
\usepackage{babel}

\newcommand{\vek}[1]{\mathbf{ #1 }}

\begin{document}

\title{Extended Ewald Summation Technique} 
\author{Ilkka Kyl\"anp\"a\"a and Esa R\"as\"anen} 
\affiliation{Department of Physics, Tampere University of Technology, P.O. Box 692, FI-33101 Tampere, Finland}
\date{\today}

\begin{abstract}
We present a technique to efficiently compute long-range interactions in systems with periodic boundary conditions. We extend the well-known Ewald method by using a linear combination of screening Gaussian charge distributions instead of only one as in the standard Ewald scheme. The combined simplicity and efficiency of our method is demonstrated, and the scheme is readily applicable to large-scale periodic simulations, classical as well as quantum.
\end{abstract}


\maketitle

In computer simulations with periodic boundary conditions the long-range potentials are usually expressed in rapidly converging sums in both real and reciprocal space according to the Ewald method of images~\cite{Ewald,AllenTildesley}. The method is in extensive use in various fields of condensed matter, material, and biological physics, to properly account for the long-range interactions, e.g., the electrostatic Coulomb interaction. In practice, however, the Ewald scheme is subject to a real-space ($r_c$) and k-space cut-offs ($k_c$), which can result in rather time-consuming computations for desired numerical accuracy.

Although improvements in the practical application of the Ewald method to various systems have been subject to extensive studies \cite{Smith,Holden}, efforts to optimize the method on the level of charge distributions are scarce. As exceptions, Natoli and Ceperley~\cite{NatoliCeperley} as well as Rajagopal and Needs \cite{Needs} have introduced different breakup schemes for the real and k-space parts, in the former work in an optimized fashion by using locally piecewise-quintic Hermite interpolant basis. The method shows significant improvement in accuracy and convergence over the standard Ewald approach, but requires additional numerical efforts in the optimization and implementation.

In this Letter, we show how to straightforwardly improve the standard Ewald summation technique. This is accomplished by using a linear combination of the screening Gaussian charge distributions instead of only one. In this way, we can modify the screening charge distribution within the same real-space cut-off to obtain a smaller k-space cut-off for the desired accuracy. Overall, our method provides a significant reduction in the computing time while maintaining a straightforward numerical implementation.

In our extended Ewald scheme the charge distribution is given by 
\begin{align}
f(r)=\sum_i c_i f_i(r),
\label{eq1}
\end{align}
where $f_i$ are Gaussian functions
\begin{align}
f_i(r)=A_{\alpha_i}e^{-\alpha_i^2 r^2}
\end{align}
with $A_{\alpha_i}=\left( \frac{\alpha_i^2}{\pi}\right)^{3/2}$ and
$r=|\vek{r}|$. The coefficients $c_i$ need to be optimized. Since
the functions $f_i$ are normalized to unity, the coefficients
are also required to add up to unity.

If all the Gaussian functions forming the screening charge
distribution have their mean value at the origin, the same
coefficients can be directly applied also to the screening potential
and the k-space coefficients. Thus, in this case, there will be an
analytical form for each term, which is convenient when calculating 
forces, for example. In practice, the summation over different terms 
is more efficient to perform once in the beginning of the simulation.

Let us consider a spherically symmetric potential $W(|\vek{r}|)$, where
$|\vek{r}|$ is the distance between two particles, and a simulation
cell of volume $\Omega$. Now the image potential is given as
\begin{align}
V(\vek{r})=Z_1Z_2 V_{\rm p}(\vek{r})=Z_1Z_2\sum_{\vek{n}} W(|\vek{r}+\vek{n}L|),
\end{align}
which includes all the interactions between a particle and the
replicas of the other particle in periodically repeated space, and
$Z_1$ and $Z_2$ are the charges of the particles. In terms of a short-range 
part (SR) and a long-range part (LR), $V_{\rm p}(\vek{r})$ can 
be written as
\begin{align}
V_{\rm p}(\vek{r})
&=
\sum_{\vek{n}} V_{\rm SR}(|\vek{r}+\vek{n}L|)+\sum_{\vek{n}} V_{\rm LR}(|\vek{r}+\vek{n}L|),
\end{align}
where 
\begin{align}
V_{\rm SR}(|\vek{r}+\vek{n}L|)
=W(|\vek{r}+\vek{n}L|)-V_{\rm LR}(|\vek{r}+\vek{n}L|).
\end{align}
Assuming that the long-range part is Fourier transformable, as it is in the case of a Gaussian charge distribution, the potential can be further modified to
\begin{align} 
V_{\rm p}(\vek{r})
&=
\sum_{\vek{n}} V_{\rm SR}(|\vek{r}+\vek{n}L|)+\sum_{\vek{k}} V_{\vek{k}}e^{i\vek{k}\cdot\vek{r}}.
\label{Eq:F1}
\end{align}
In a more practical form, i.e., in the presence of a neutralizing
background, or under the assumption of charge neutrality, we can write
\begin{align}
V_{\rm p}(\vek{r})
&=
\sum_{\vek{n}} V_{\rm SR}(|\vek{r}+\vek{n}L|)+\sum_{\vek{k}\neq \vek{0}} V_{\vek{k}}e^{i\vek{k}\cdot\vek{r}}+C_V,
\end{align}
where
\begin{align}
C_V=-\frac{1}{\Omega}\int d\vek{r} \left[W(\vek{r})-V_{\rm LR}(\vek{r})\right].
\nonumber
\end{align}
This term represents contributions from a neutralizing background, and
in the total energy these $C_V$ terms will cancel out for charge
neutral systems.  In the calculation of the total energy the Madelung
constant ($V_{\rm M}=\tfrac{1}{2}\lim_{r\rightarrow 0}[V_{\rm
    p}(r)-W(r)]$) is also needed, the energy term being $\sum_i Z_i^2
V_{\rm M}$.

In the case of a single screening Gaussian term, 
the long-range potential, or screening potential, is given in 
real space as
\begin{align}
V_{\rm LR}(r)=\frac{{\rm erf}(\alpha r)}{r},
\label{Eq:Vlr_s}
\end{align}
and in reciprocal space the Fourier coefficients are
\begin{align}
V_{\vek{k}}=V_k=\frac{4\pi}{\Omega}\frac{e^{-k^2/4\alpha^2}}{k^2}.
\label{Eq:sEwVk}
\end{align}
As mentioned earlier, in the case of a linear combination of screening
Gaussian charge distributions with zero mean, the potentials are given as 
\begin{align}
V_{\rm LR}(r)=\sum_i c_i V_{{\rm LR},i}(r)=\sum_i c_i \frac{{\rm erf}(\alpha_i r)}{r},
\label{Eq:Vlr_e}
\end{align}
and
\begin{align}
V_k=\sum_i c_i V_{k,i}(r)=\frac{4\pi }{\Omega} \sum_i c_i \frac{e^{-k^2/4\alpha_i^2}}{k^2},
\label{Eq:eEwVk}
\end{align}
where index $i$ refers to different $\alpha$-parameters,
i.e., Gaussian functions with different variances.

At this point we have introduced all the equations that are needed in
performing either the standard Ewald summation or the extended Ewald
technique. In theory, both cases are {\em exact} in the presented form
for any given set of $\alpha$-parameters and $c_i$ coefficients (for
which $\sum_i c_i=1$). In practice, however, it is critical to choose
proper values for the cut-offs $r_c$ and $k_c$ in order to obtain good
accuracy and reasonable computation time.

Here we choose the real-space cut-off to be $r_c=L/2$, which
restricts the potential to be a function of the minimum distance of
a particle to any image. Thus, the set of $\alpha$-parameters needs to
ensure that 
\begin{align}
\sum_{\vek{n}\neq \vek{0}} V_{\rm SR}(|\vek{r}+\vek{n}L|)\approx 0,
\label{Eq:MIapprox}
\end{align}
that is, $W(|\vek{r}|)\approx V_{\rm LR}(|\vek{r}|)$ for all $r>r_c$.
Now the image potential of Eq.~\eqref{Eq:F1} can be written as
\begin{align} 
V_{\rm p}(\vek{r})
&\approx
V_{\rm SR}(r)\Theta(r_c-r)+\sum_{|\vek{k}|\le k_c} V_{k}e^{i\vek{k}\cdot\vek{r}}
+\sum_{|\vek{k}|>k_c} V_{k}e^{i\vek{k}\cdot\vek{r}},
\label{Eq:Iapprox}
\end{align}
where we also included the cut-off in the reciprocal space, and $\Theta(x)$
is the Heaviside step function. If the short-range part is truncated accurately
according to Eq.~\eqref{Eq:MIapprox}, the k-space cut-off will be
the source for the accuracy in the image potential. Therefore, if
$\Delta$ represents the error due to the k-space cut-off, we have
\begin{align}
\Delta & = 
\sum_{|\vek{k}|>k_c} V_{k}e^{i\vek{k}\cdot\vek{r}}.
\end{align}
Here it should be pointed out that for the optimized breakup of Natoli
and Ceperley~\cite{NatoliCeperley} the constraint of
Eq.~\eqref{Eq:MIapprox} is exact, since the screening charge
distribution is equal to zero for $r\ge r_c$. For Gaussian
distributions this will become exact only in the limit 
$\alpha\rightarrow\infty$, but highly accurate approximations 
can be made with reasonable values of $\alpha$.

In the case of the conventional Ewald method, if the condition in
Eq.~\eqref{Eq:MIapprox} is fulfilled accurately, the k-space part will
usually end up being slowly convergent, and the convergence is solely
determined by the Gaussian parameter $\alpha$, see
Eq.~\eqref{Eq:sEwVk}. However, in the case of multiple Gaussian
distributions, the $V_k$ is given by Eq.~\eqref{Eq:eEwVk}, and thus,
we have
\begin{align}
\Delta & = 
\sum_{|\vek{k}|>k_c} \sum_{n}c_n V_{k,n}e^{i\vek{k}\cdot\vek{r}}.
\label{Eq:Delta}
\end{align}
Now the convergence in k-space is affected by the coefficients $c_i$
as well as the Gaussian parameters $\alpha_i$. Since
$V_{\vek{k}}=V_{-\vek{k}}=V_k$, the above expression can be written in terms of 
a cosine function, and also the summation order can be changed:
\begin{align}
\Delta & = 
\sum_{n} c_n \sum_{|\vek{k}|>k_c} V_{k,n}\cos(\vek{k}\cdot\vek{r}).
\end{align}
This expression can be already used to obtain the coefficients $c_i$
for a predefined set of $\alpha_i$-parameters. First, we define a 
three-dimensional grid, $x=-L/2 \ldots L/2$, $y=-L/2 \ldots L/2$ and $z=-L/2
\ldots L/2$. Secondly, we construct a matrix equation from the equation
above and use also the fact that $\sum c_i=1$. Solving this
over-determined set of linear equations results in the least-squares
solution for the integrand in
\begin{align}
\chi^2=\frac{1}{\Omega}\int_{\Omega} d\vek{r}\Delta^2.
\label{Eq:chi2}
\end{align}
Thirdly, we can compute the integral with the obtained coefficients in order to
estimate the $\chi^2$ error.

Another, improved way~\cite{NatoliCeperley} to achieve the
coefficients $c_i$ is to start from Eqs.~\eqref{Eq:Delta} and
\eqref{Eq:chi2}, i.e.,
\begin{align}
\chi^2 &= \frac{1}{\Omega}\int_{\Omega} d\vek{r}\left( \sum_{|\vek{k}|>k_c} \sum_{n}c_n V_{k,n}e^{i\vek{k}\cdot\vek{r}} \right)^2
\nonumber\\
&=
\frac{1}{\Omega}\int_{\Omega} d\vek{r}
\sum_{|\vek{k}|>k_c}\sum_{|\vek{k}'|>k_c} \sum_{n}\sum_{m}c_n c_m V_{k,n}V_{k',m}
e^{i(\vek{k}+\vek{k}')\cdot\vek{r}}
\nonumber\\
&=
\sum_{|\vek{k}|>k_c}\sum_{|\vek{k}'|>k_c} \sum_{n}\sum_{m}c_n c_m V_{k,n}V_{k',m}
\delta_{\vek{k},-\vek{k}'}
\nonumber\\
&=
\sum_{|\vek{k}|>k_c}\sum_{n}\sum_{m}c_n c_m V_{k,n}V_{k,m}.
\end{align}
Next, let us take the derivative of this expression 
with respect to $c_n$ and set it be equal to zero, that is,
\begin{align}
\frac{\partial \chi^2}{\partial c_n} = 0,
\end{align}
which for each $|\vek{k}|>k_c$ leads to
\begin{align}
\sum_{m} c_m V_{k,m} = 0.
\end{align}
For each $\vek{k}$ we have a linear equation, and therefore, together
with the constraint $\sum c_i=1$, we have an over-determined set of linear
equations, which minimizes the $\chi^2$. After having determined the
coefficients, $\chi^2$ can be computed from
\begin{align}
\chi^2=\sum_{|\vek{k}|>k_c}(\sum_i c_i V_{k,i})^2.
\end{align}

\begin{figure}[t]
\includegraphics[width=8cm]{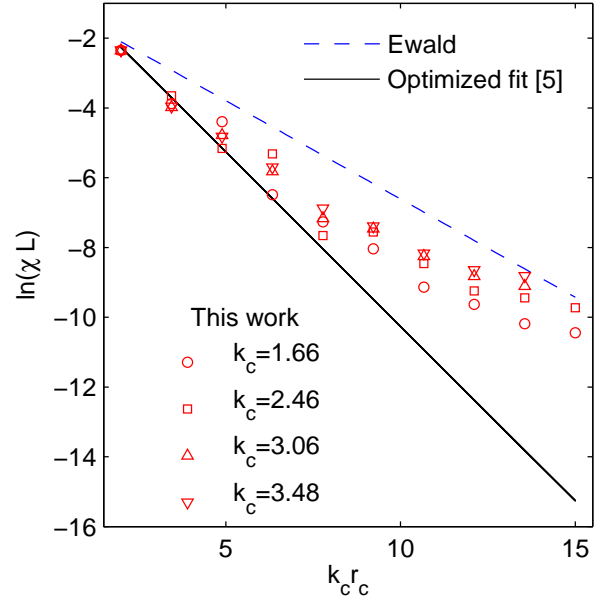}
\caption{\label{fig1} (Color online) Natural logarithm of $\widetilde{\chi}L$ for the Coulomb potential as a function of the dimensionless parameter
  $k_cr_c$ for our extended Ewald scheme (red symbols), the optimized break-up by Natoli and Ceperley~\cite{NatoliCeperley} (solid line), and the optimized standard Ewald method (dashed line).}
\end{figure}

In finding the coefficients $c_i$ for the Gaussian functions above we assumed 
that Eq.~\eqref{Eq:MIapprox} holds accurately. This is a valid assumption 
for sufficiently large values of $\alpha$. However, it restricts the degrees of freedom in the optimization, which can be released by a new term
\begin{align}
\widetilde{\Delta}=\sum_{|\vek{k}|>k_c} V_{k}e^{i\vek{k}\cdot\vek{r}}+\sum_{\vek{k}}\widetilde{V_k}e^{i\vek{k}\cdot\vek{r}},
\end{align}
where 
\begin{align}
\widetilde{V_k}=\frac{1}{\Omega}\int d\vek{r} V_{\rm SR}(r)[1-\Theta(r_c-r)]e^{i\vek{k}\cdot \vek{r}}.
\end{align}
In this work the potentials were chosen to be spherically symmetric,
and therefore, $\widetilde{V_k}$ can also be written as
\begin{align}
\widetilde{V_k}=\frac{4\pi}{\Omega k}\int_{0}^{\infty} dr ~rV_{\rm SR}(r)[1-\Theta(r_c-r)]\sin(kr).
\end{align}
With this $\widetilde{\Delta}$ term the exact equality in
Eq.~\eqref{Eq:Iapprox} is restored, i.e.,
\begin{align} 
V_{\rm p}(\vek{r})
=
V_{\rm SR}(r)\Theta(r_c-r)+\sum_{|\vek{k}|\le k_c} V_{k}e^{i\vek{k}\cdot\vek{r}}
+\widetilde{\Delta}.
\end{align}

Now, $\widetilde{\chi}^2$ may be expressed as
\begin{align}
\widetilde{\chi}^2 & =\frac{1}{\Omega}\int_{\Omega} d\vek{r} \widetilde{\Delta}^2
\nonumber\\
&=
\sum_{\vek{k}}\left( A_k + \sum_n c_n B_{k,n} \right)
\left( A_k + \sum_m c_m B_{k,m} \right),
\end{align}
where $A_k$ and $B_{k,i}$ are defined as
\begin{align}
A_k &=W_k-\frac{4\pi}{\Omega k}\int_{0}^{r_c} dr~r W(r)\sin(kr),
\\
B_{k,i} &=\frac{4\pi}{\Omega k}\int_{0}^{r_c} dr~r V_{{\rm LR},i}(r)\sin(kr)
-V_{k,i}\Theta(k_c-k),
\end{align}
where $W_k$ is the Fourier coefficient of $W(|\vek{r}|)$. Setting the
derivative of $\widetilde{\chi}^2$ with respect to $c_n$ to zero leads to 
a set of linear equations for each $\vek{k}$:
\begin{align}
\sum_{m} c_m B_{k,m} = -A_k,
\label{Eq:optim}
\end{align}
which for $\vek{k}=\vek{0}$ reduces to $\sum_m c_m=1$, and the
accuracy can be estimated by
\begin{align}
\widetilde{\chi}^2=\sum_{\vek{k}} \left( A_k + \sum_i c_i B_{k,i}\right)^2.
\end{align}

\begin{figure}[t]
\includegraphics[width=7cm]{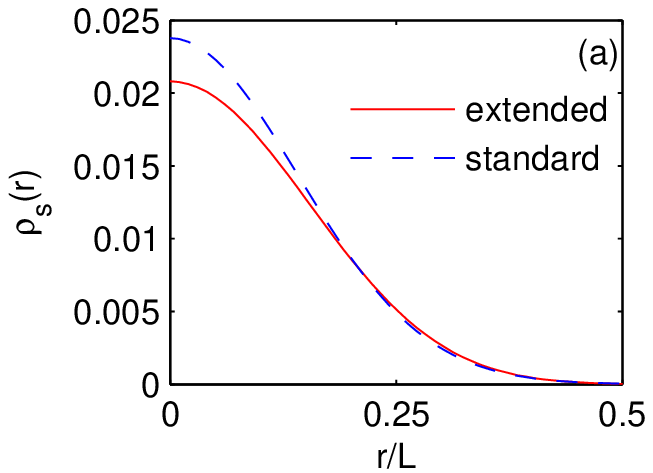}
\includegraphics[width=7cm]{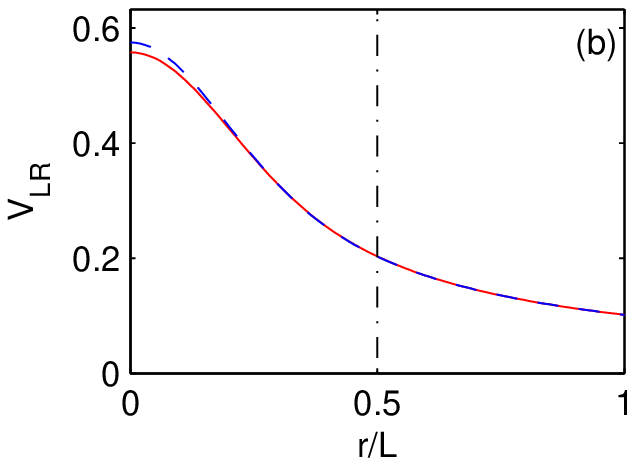}
\includegraphics[width=7cm]{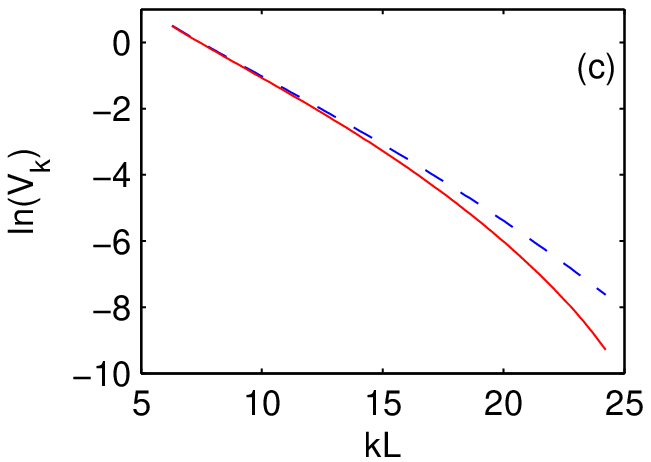}
\caption{\label{fig2} (Color online) (a) Screening charge distributions for
  $k_cr_c\approx 12.11$ calculated with our extended Ewald scheme 
(solid line) and the standard scheme (dashed line). 
(b) Corresponding long-range potentials, see also
  Eqs.~\eqref{Eq:Vlr_s} and \eqref{Eq:Vlr_e}. Cut-off $r_c$ is given as
  a vertical dash-dotted line. (c) Natural logarithm of the Fourier
  coefficients.}
\end{figure}

As an example, let us consider the commonly-used Coulomb potential with the 
standard Ewald method in comparison with our extended scheme. The
potential is given by $W(r)=1/r$ and its Fourier coefficients are $W_k=4\pi/\Omega k^2$. The product of charges, $Z_1Z_2$, is set equal to one. 
Here we use Newton's method for finding a minimum for the function
$g(\{c_i\},\{\alpha_i\})=\widetilde{\chi}^2+\lambda\left( \sum_{i}c_i - 1 \right)^2$, where $\lambda$ is a Lagrange multiplier.

Fig.~\ref{fig1} shows the error $\ln\left(\widetilde{\chi} L\right)$ as
a function of $k_cr_c$ for both the conventional Ewald method (dashed line) and our extended scheme for various $k_c$ and $r_c$ values (symbols), together with the the fit by Natoli and Ceperley~\cite{NatoliCeperley} (solid line).
Remarkably, the extended Ewald approach improves the accuracy by more
than an order of magnitude over the conventional Ewald case.
On the other hand, the ``optimized break-up''~\cite{NatoliCeperley} acts as a
lower bound estimate for our values. It should be pointed out that the results of our scheme can be improved further by enhanced optimization. This could involve a combination of the three schemes introduced here along with optimizing the Gaussian $\alpha_i$-parameters, for example. 

The screening charge distribution for $k_cr_c\approx 12.11$ is shown
in Fig.~\ref{fig2}(a). For the extended Ewald scheme the distribution is
spread out more than in the standard Ewald case. Both
distributions converge close to zero before the real-space cut-off
$r_c$. With these optimized coefficients the effect seems to be more pronounced here than in the optimized break-up case, see Fig.~4 in Ref.~\cite{NatoliCeperley}.
 
In Fig.~\ref{fig2}(b) we show the long-range potentials of
Eqs.~\eqref{Eq:Vlr_s} and \eqref{Eq:Vlr_e} corresponding to the
distributions shown in Fig.~\ref{fig2}(a). The potentials are different
from origin to roughly $0.3L$, after which (in the scale of the figure)
the potentials coincide well before the real-space cut-off ($r_c=L/2$). 
In the differing range the changes in the potential are smoother in the 
extended scheme, and thus, the Fourier coefficients converge considerably 
faster than in the standard approach, which is demonstrated 
in Fig.~\ref{fig2}(c).

In addition to the improved accuracy, another clear advantage of the
extended Ewald scheme is the fact that it is easily adaptable to numerical codes already having the standard Ewald method. Moreover, in the extended scheme the
analytical form is preserved, which is advantageous when calculating
accurate derivatives of the potentials to obtain forces, for example.
It is also important to note that, regardless of the number of terms in the extended scheme, computations will not be more time consuming, since in any case
a radial potential (with an error function) should be interpolated
from a radial grid during the simulation. Therefore, the linear
combination coefficients are needed only in the beginning of the
simulation.

In this Letter we have demonstrated that a linear combination of
Gaussian functions as the screening charge distribution can be used to
considerably improve the standard Ewald method of images.  The
modified charge distribution enables smaller reciprocal space cut-off
than only a single Gaussian function for a higher level of
accuracy. The extended scheme leads to reduced computer time in
simulations of periodic systems and it can be easily implemented in
any numerical package using periodic boundary conditions within, e.g.,
density-functional methods, molecular dynamics, and classical or
quantum Monte Carlo calculations.  The full potential of the present
technique can be achieved by a further developed optimization
procedure.

We thank Jouko Nieminen and Tapio Rantala for useful discussions. This work was supported by the Academy of Finland, COST Action CM1204 (XLIC), Nordic Innovation through its Top-Level Research Initiative (project no. P-13053), and the European Community's FP7 through the CRONOS project, grant agreement no. 280879.


\end{document}